%Paper: hep-th/9211002
%From: 402g0027@ex.ecip.osaka-u.ac.jp
%Date: Mon, 2 Nov 92 18:39:17 +0900

%%
%%   Isse  1992/10/1 11:00
%%
%%%%%%%%%%%%%%%%%%%%%%%%%%%%
\input phyzzx
\rightline{OU-HET 170}
%\publevel{preprint}
\rightline{November, 1992}
\vskip2.0cm
\centerline{\bf Unification of Gravity, Gauge and Higgs Fields}
\centerline{\bf by Confined Quantum Fields}
\centerline{\bf -Mathematical Formulation-}
\vskip1.5cm
\centerline{Toshiki Isse\footnote*{402g0027@ex.ecip.osaka-u.ac.jp}}
%%%%%%%%%%%%%%%%%%%%%%%%%%%%%%%%%%%%%%%%%%%

\vskip1.5cm

\centerline {\it Department of Physics}
\centerline {\it Osaka University, Toyonaka, Osaka 560, Japan}

\vskip2cm

  Dynamics of quantized free fields ( of spin 0 and 1/2 ) contained in a
subspace $V_*$ of an N+4 dimensional flat space $V$ is studied. The space $V_*$
is considered as a neighborhood of a four dimensional submanifold $M$
arbitrarily embedded into $V$. We study the system as a simple model of unified
theory of gravity ($g$), SO(N) gauge fields ($A$) and Higgs fields ($\phi $).
In this paper classical treatment of the system is given. We show that,
especially when the fields have spin 1/2, the system is described by an
infinite number of fields in $M$ interacting with $g$, $A$ and $\phi $. The
fields $g$, $A$ and $\phi $ are induced themselves by embedding functions of
$M$ and correspond respectively to induced metric, normal connection and
extrinsic curvature of $M$.
\medskip

\vfill\eject

%%%%%%%%%%%%%%%%%%%%%%%%%%%%%%%%%%%%%%%%%%%%%%%%%%%%%%%%%%%%%%%%%%%
%%%%%%%%%%%%%%%%%%%%%%%%%%%%%%%%%%%%%%%%%%%%%%%%%%
 \beginsection 1 Introduction

 Bulk properties of quantized fields are probed by a response of the vacuum
state to constraints imposed upon the system. Well-studied examples include the
casimir effect [1], which is characterized by mechanical forces due to
quantized fields in a finite region. These mechanical properties of the
quantized fields have been discussed by many authors [2][3].
  In this and the next papers [4] we discuss dynamics of quantized fields
contained in a subspace $V_*$ of N+4 dimensional flat space $V$. We regard
$V_*$ as a neighborhood of a four dimensional submanifold $M$ arbitrarily
embedded into $V$.( The exact definition of the neighborhood will be given in
section three. ) In other words, we assume an existence of some physical
object\footnote*{The reader can easily understand the situation if he imagines
a tube-like object which is homeomorphic to $S^{N-1} \times M$ and which is
embedded ioto $V$.} (we call it $B%a(Jtemplate$B%b(J of spacetime) which
%%confines quantized fields inside the neighborhood of $M$. We study the
%%dynamics of the quantized fields confined in the object, see Fig[1] .
 To avoid undesirable complexities, we assume also that the
$B%a(Jtemplate$B%b(J of spacetime is so flexible that it has no mechanical
%%properties by itself. If the radius $\ell /2$ of the tube-like object in
%%Fig[1] is small enough, we can consider the region inside it as four
%%dimensional spacetime approximately. Investigating the response of the vacuum
%%of the system to how the submanifold $M$ is embedded into $V$, we show that
%%Einstein SO(N)-Yang-Mills-Higgs theory is induced as a low energy effective
%%theory of the system[4]. Gravity, gauge fields and Higgs fields are induced
%%by an embedding $f$ of $M$ into $V$, which is specified by  N+4 embedding
%%functions;
$$
Z^\alpha =X^\alpha (x_\mu )\  ( \ \mu =0,1,2, 3\ \alpha =0,1,...,N +3\
).\eqno(1.1)
$$
 Here $x_\mu (\mu =0,1,2,3)$ are coordinates of $M$ and $Z^\alpha (\  \alpha
=0,1,...,N+3 \ )$ are Cartesian coordinates of $V$. In the language of
differential geometry, the gravity, the gauge fields and the Higgs fields
introduced above are identified respectively as the induced metric, the normal
connection and the trace of extrinsic curvature of $M$ [5]. These fields are
determined by how the spacetime manifold is embedded. Therefore the system can
be considered as a simple model of unified theory of gravity, SO(N)-gauge  and
Higgs fields.
   In section two, as a warming up example, we consider classical dynamics of
free scalar field contained in a neighborhood of a circle embedded into a two
dimensional Euclidean space. In section three, we discuss classical dynamics of
a free scalar field. This field is contained in a neighborhood of the four
dimensional submanifold $M$ which is embedded into the N+4 dimensional flat
space. In section four, the spinor case will be discussed. The classical
dynamics of the system, discussed in section  four, is described by an infinite
number of fields in $M$ interacting with gravity and SO(N) gauge fields. In
section five, we show that there exists a total derivative term ( in $V$ )
which induces Higgs fields and a Yukawa interaction term. It should be noted
that the theory considered here is different from Kaluza-Klein type theory
[6][7] in many respects. We comment on this point in section six.

\beginsection 2  A simple example  in two dimension

 Consider a free scalar field which is contained in a tubular neighborhood
$V_*$ of a circle of radius $a$ (which should be denoted by $M$ ). The circle
is embedded into a two dimensional Euclidean space $E_2$ ( $E_2=V$). Here the
tubular neighborhood of the embedded submanifold $M$ is a set of all points in
$V$ whose distance from $M$ is less than ${l/2\ (l<<a)}$ Fig[2].

FIG[1]

 We first describe the action of the free scalar field $D(Z)$ in two
dimensional Euclidean space $E_2$;
$$
S[D]=\int_{E^2} {dZ^2}\sum\limits_{\alpha =0}^1 {{{\partial D^*(Z)} \over
{\partial Z^\alpha }}{{\partial D(Z)} \over {\partial Z^\alpha }}}.\eqno(2.1)
$$
 Here $(Z^0,Z^1)$ are Cartesian coordinates of $E_2$. The summations over the
indices is suppressed from now on. If the field is confined into the tubular
neighborhood of  $S^1$, the action is replaced by
$$
S[D,S^1]=\int_{V_*} {dZ^2}{{\partial D^*(Z)} \over {\partial Z^\alpha
}}{{\partial D(Z)} \over {\partial Z^\alpha }},\eqno(2.2)
$$
 Here $V_*$ stands for the neighborhood of $S^1$and we impose the Neumann
boundary condition\footnote*{ The Dirichlet boundary condition is inadequate
here; it forbids zero modes to appear alone in the action (2.9).} on the
boundary of $V_*$. The embedding functions $X^\alpha (x^0)$ of $S^1$are given
as follows;
$$
X^0(x^0)=a\sin  (x^0/a),\eqno(2.3a)
$$
$$
X^1(x^0)=a\cos  (x^0/a), \ (0\le x^0<2\pi a).\eqno(2.3b)
$$
 We perform a coordinate transformation in the neighborhood of $S^1$ from
$(Z^0,Z^1)$ to $(x^0,\varsigma )$,
$$
Z^0=(a+\varsigma )\sin  (x^0/a),\eqno(2.4a)
$$
$$
Z^1=(a+\varsigma )\cos  (x^0/a), \ (0\le x^0<2\pi a ,\ -l/2\le \varsigma \le
l/2).\eqno(2.4b)
$$
 In the new coordinates, the line element is given by,
$$
ds^2=g_{00}(x^0,\varsigma )dx^0dx^0+d\varsigma d\varsigma ,\eqno(2.5)
$$
where $g_{00}(x^0,\varsigma )=(a+\varsigma )^2/a^2$. The action is given by
$$
\eqalign{&S[D,S^1]=\int\limits_0^{2\pi a} {dx^0\sqrt
{g_{00}(x^0)}}\int\limits_{-l/2}^{l/2} {d\varsigma }[g^{00}(x^0)(1-3{\varsigma
\over a}+0[({\varsigma  \over a})^2]){{\partial D^*} \over {\partial
x^0}}{{\partial D} \over {\partial x^0}}\cr
  &+(1-{\varsigma  \over a}+0[({\varsigma  \over a})^2]){{\partial D^*} \over
{\partial \varsigma }}{{\partial D} \over {\partial \varsigma
}}],\cr}\eqno(2.6)
$$
where $g_{00}(x^0)=1$ is an induced metric of $S^1$ from $V$.
 Hereafter we neglect  $0({\varsigma  \over a})$ terms, as we are interested
only in the limit ${l/a\to 0}$.
  We decompose the field $D$ as a Fourier series with respect to $\varsigma $,
$$
D(x^0,\varsigma )=l^{-1/2}\sum\limits_{n=-\infty }^{+\infty } {D(x^0,n)\exp
(i2\pi n\varsigma /l)}.\eqno(2.7)
$$
 The Neumann boundary condition is
$$
\left. {{{\partial D(x^0,\varsigma )} \over {\partial \varsigma }}}
\right|_{\varsigma =\pm l/2}=0.\eqno(2.8)
$$	We finally obtain the following action by inserting the expression (2.7)
into eq. (2.6),
$$\eqalign{&S[D,S^1]\cr
  &=\int\limits_0^{2\pi } {dx\sqrt {g_{00}(x)}\sum\limits_{n=-\infty }^\infty
{(-g^{00}(x){{\partial D^*(x,n)} \over {\partial x}}{{\partial D(x,n)} \over
{\partial x}}+(2\pi n/l)^2\left| {D(x,n)} \right|^2})}.\cr}\eqno(2.9)
$$
   We have neglected $O({\varsigma  \over a})$terms and an integration with
respect to $\varsigma $ is performed. Eq.(2.9) is given in one dimensional
spacetime $S^1$.

\beginsection 3 N+4 dimensional case- scalar field

In this section we discuss classical dynamics of a free massless scalar field
that is contained in a neighborhood $V_*$ of $M$. We denote by $M$  a four
dimensional submanifold arbitrarily embedded into N+4 dimensional flat space
$V$. The metric of $V$ is $\eta _{\alpha \beta }\dot =(-1,1,...,1)$  $(\alpha
,\beta =0,1,...,N+3)$.
  As in the previous section we first consider free field theory of the
massless scalar field $D$ in $V$. This system is described by the action,
$$
S_{scalar}[D]=\int\limits_V^{} {dZ^{N+4}(-\eta ^{\alpha \beta }{{\partial
D^*(Z)} \over {\partial Z^\alpha }}{{\partial D(Z)} \over {\partial Z^\beta
}})},\eqno(3.1)
$$
where $Z^\alpha (\alpha =0,1,...,N+3)$ are Cartesian coordinates of $V$. If the
scalar field is constrained in a neighborhood of $M$ whose embedding is
specified by the N+4 functions $X^\alpha (x^\mu )$, the action becomes,
$$
S_{scalar}[D,X]=\int\limits_{V_*}^{} {dZ^{N+4}(-\eta ^{\alpha \beta }{{\partial
D^*(Z)} \over {\partial Z^\alpha }}{{\partial D(Z)} \over {\partial Z^\beta
}})},\eqno(3.2)
$$
where $V_*$ stands for the neighborhood of $M$. The exact definition of the
neighborhood will be given shortly. An appropriate coordinate system to
describe $V_*$ is given by $(x_\mu ,\varsigma ^a)$ where $x_\mu (\mu =0,1,2,3)$
are the coordinates tangent to $M$ and $\varsigma ^a(\  a=1,...,N)$ are those
normal to $M$. See Fig[3]. The line element of $V$ is written in the new
coordinates,
$$
ds^2=\sum\limits_{\mu ,\nu =0}^3 {g_{\mu \nu }(x,\varsigma )dx^\mu dx^\nu
}+\sum\limits_{a=1}^N {d\varsigma ^ad\varsigma ^a}.\eqno(3.3)
$$
 Furthermore we can choose $\varsigma ^a=0$ on the manifold $M$, so that $x_\mu
(\mu =0,1,2,3)$ can be regarded as the coordinates of $M$. Using the new
coordinates, we define the neighborhood $V_*$ as a region\footnote*{This
definition of the neighborhood is adopted for analytical reason. It might be
more natural to adopt the tubular neighborhood as $V_*$.} that satisfies
$\left| {\varsigma ^a} \right|\le l/2,\  (a=1,...,N)$. We can always find the
above coordinates if all focal points lie outside the neighborhood of $M$
Fig[3]. ( A focal point of $M$ is a point of $V$ where nearby normals intersect
[13].) In order to satisfy the above condition, $M$ has to be embedded so that
the smallest focal length $d(X)$ is larger than $\ell /2$. In the following
section we will introduce the second fundamental tensor (or extrinsic
curvature) of embedded spacetime $M$ to evaluate the focal length.

Fig[3]

 The new coordinates in the neighborhood of M and a focal point.In the new
coordinates, the Lagrangian of the scalar field is written by
$$
\eqalign{&S_{scalar}[D,X]=\int\limits_{V_*}^{} {\sqrt {-g(X(x))}dx^4d\varsigma
^N\ [g^{\mu \nu }(X(x))(1+\Delta ){{\partial D^*(x,\varsigma )} \over {\partial
x^\mu }}{{\partial D(x,\varsigma )} \over {\partial x^\nu }}}\cr
  &\                     \ \ \ \ \ \ \ \ \ \ \ \ \ \ \ +(1+\Delta '){{\partial
D^*(x,\varsigma )} \over {\partial \varsigma ^b}}{{\partial D(x,\varsigma )}
\over {\partial \varsigma ^b}}],\cr}\eqno(3.4)
  $$
where $g_{\mu \nu }(X(x))$ is the induced metric of $M$ ,
$$
g_{\mu \nu }(X(x))={{\partial X^\alpha } \over {\partial x^\mu }}{{\partial
X_\alpha } \over {\partial x^\nu }},\eqno(3.5)
$$
and $\Delta $and $\Delta '$ are respectively defined as ,
$$
\eqalign{&\Delta =\varsigma ^a(-{{g^{\rho \sigma },_a(x)g_{\rho \sigma }(x)}
\over 2}+{{g^{\mu \nu },_a(x)} \over {g^{\mu \nu }(x)}})\cr
  &\            +{1 \over 2}\varsigma ^a\varsigma ^b(-{{g^{\rho \sigma
}(x)g_{\rho \sigma ,ab}(x)} \over 2}-{{g^{\rho \sigma }(x)g_{\rho \sigma
,a}(x)g^{\mu \nu },_b(x)} \over {g^{\mu \nu }(x)}}+{{g^{\mu \nu },_{ab}(x)}
\over {g^{\mu \nu }(x)}})+\cdot \cdot \cdot ,\cr}
$$
$$
\Delta '=\varsigma ^a(-{{g^{\rho \sigma },_a(x)g_{\rho \sigma }(x)} \over
2})+{1 \over 2}\varsigma ^a\varsigma ^b(-{{g^{\rho \sigma }(x)g_{\rho \sigma
,ab}(x)} \over 2})+\cdot \cdot \cdot .\eqno(3.6)
$$
( The notations $\Delta $, $\Delta '$ make sense only in the action (3.4). )
The fields $g^{\rho \sigma },_a(x)$ and $g^{\rho \sigma },_{ab}(x)$are
respectively given by
$$
g^{\rho \sigma },_a(x)\equiv \left. {{{\partial g^{\rho \sigma }(x,\varsigma )}
\over {\partial \varsigma ^a}}} \right|_{\varsigma ^a=0},\eqno(3.7)
$$
$$
g^{\rho \sigma },_{ab}(x)\equiv \left. {{{\partial ^2g^{\rho \sigma
}(x,\varsigma )} \over {\partial \varsigma ^a\partial \varsigma ^b}}}
\right|_{\varsigma =0}.\eqno(3.8)
$$
 The Neumann boundary condition is imposed at the boundary of $V_*$,
$$
{{\partial D(x^\mu ,\varsigma )} \over {\partial \varsigma ^a}}=0.\eqno(3.9)
$$
 Using dimensional analysis and taking into account the fact that the focal
length determines the characteristic scale of the manifold $M$, we find
$$
\Delta \cong \Delta '\cong O({\ell  \over {d(X)}}).\eqno(3.10)
$$
 As in the previous section, we neglect the correction terms $\Delta $and
$\Delta '$ as we are interested only in the limit ${l/d(X)\to 0}$ ( we will
discuss the physical meaning of the limit in the next paper ). Then we
decompose the scalar field as a Fourier series with respect to $\varsigma ^a$,
$$
D(x^\mu ,\varsigma ^a)=l^{-N/2}\sum\limits_n^{} {D(x^\mu ,n)\exp (i2\pi
n^a\cdot \varsigma ^a/l)}.\eqno(3.11)
$$
 We substitute the above expression into the action (3.4) and integrate over
the $\varsigma ^a$ variables, neglecting $\Delta $and $\Delta '$. We find
$$
\eqalign{&S_{scalar}[D,X]\cr
  &=\int\limits_M^{} {dx^4\sqrt {g(x)}\sum\limits_n^{} {(-g^{\mu \nu
}(x){{\partial D^*(x,n)} \over {\partial x^\mu }}{{\partial D(x,n)} \over
{\partial x^\nu }}+(2\pi n/l)^2D^*(x,n)D(x,n))}}\cr}\eqno(3.12)
$$
 The dynamics of the system obtained as the action of four dimensional
spacetime is nothing but an infinite number of scalar fields in curved four
dimensional space. Different modes would couple with each other in the action
(3.12) if the correction terms $\Delta $and $\Delta '$are present. The system
depends only on the intrinsic geometry of $M$. If we consider spinor fields
confined in a neighborhood of $M$ however, the situation is different. We
cannot describe the corresponding system, consisting of the spinor fields, by
the intrinsic geometry alone. The spin of the fields feels a relation between
$V$ and the embedded manifold $M$. Such information is given by the SO(N) gauge
fields and the extrinsic curvature defined in the next section.

\beginsection 4   N+4 dimensional case -  spinor fields

Let us first consider free field theory of massless spinor fields $\Phi $ with
$2^{[(N+4)/2]}$ components in $V$. It is described by the action
$$
S_{spinor}[\Phi ]={1 \over 2}\int\limits_{\  V}^{} {dZ^{N+4}i\bar \Phi
(Z)\Gamma ^\alpha ({{\vec \partial } \over {\partial Z^\alpha
}}-{{\mathord{\buildrel{\lower3pt\hbox{$\scriptscriptstyle\leftarrow$}}\over
\partial } } \over {\partial Z^\alpha }})\Phi (Z)},\eqno(4.1)
$$
where $\Gamma ^\alpha $ (a=0,1,...,N+3 ) satisfy a Clifford algebra
$$
{\Gamma _\alpha ,\Gamma _\beta }=2\eta _{\alpha \beta }.\eqno(4.2)
$$
 Using $2^{[N/2]}\times 2^{[N/2]}$ matrices $\hat \Gamma ^a$ ${a=1,...,N}$
which satisfy
$$
{\hat \Gamma ^a,\hat \Gamma ^b}=2\delta ^{ab},\eqno(4.3)
$$
 We write $\Gamma ^\alpha $ as direct products of  $\hat \Gamma ^a$ and the
Dirac matrices $\gamma ^l$,
$$
\Gamma ^l=\gamma ^l\otimes 1\  (l=0,...,3),\eqno(4.4a)
$$
$$
\Gamma ^{3+a}=\gamma ^5\otimes \hat \Gamma ^a\  (a=1,...,N).\eqno(4.4b)
$$
 The action of the spinor fields in $V_*$ becomes,
$$
S_{spinor}[\Phi ,X]={1 \over 2}\int\limits_{V_*}^{} {dZ^{N+4}i\bar \Phi
(Z)\Gamma ^\alpha ({{\vec \partial } \over {\partial Z^\alpha
}}-{{\mathord{\buildrel{\lower3pt\hbox{$\scriptscriptstyle\leftarrow$}}\over
\partial } } \over {\partial Z^\alpha }})\Phi (Z)}.\eqno(4.5)
$$
 As in the scalar field case, we impose the Neumann boundary condition at the
boundary of $V_*$ and we write the eq.(4.5) in the coordinates $(x_\mu
,\varsigma ^a)$ $(\  \mu =0,1,2,3 \ a=1,...,N)$, neglecting $O(\ell /d(X))$
corrections;
$$
\eqalign{&S_{spinor}[\Phi ',X]={1 \over 2}\int\limits_{\  V_*}^{} {\sqrt
{-g(x)}dx^4d\varsigma ^N[i\bar \Phi '(x,\varsigma )\Gamma ^\mu (x){{\partial
\Phi '(x,\varsigma )} \over {\partial x^\mu }}i\bar \Phi '(x,\varsigma )+}\cr
  &-{{\partial i\bar \Phi '(x,\varsigma )} \over {\partial x^\mu }}\Gamma ^\mu
(x)\Phi '(x,\varsigma )+\left. {\bar \Phi '(x,\varsigma )\Gamma ^\alpha
{{\partial \varsigma ^a} \over {\partial Z^\alpha }}} \right|_{\varsigma
=0}{{\partial \Phi '(x,\varsigma )} \over {\partial \varsigma ^a}}\cr
  &-{{\partial \bar \Phi '(x,\varsigma )} \over {\partial \varsigma ^a}}\left.
{\Gamma ^\alpha {{\partial \varsigma ^a} \over {\partial Z^\alpha }}}
\right|_{\varsigma =0}\Phi '(x,\varsigma )],\cr}\eqno(4.6)
  $$
where
$$
\Phi '(x^\mu ,\varsigma ^a)=\Phi (Z^\alpha (x^\mu ,\varsigma ^a)),\eqno(4.7)$$
$$
\Gamma ^\mu (x)\equiv \sum\limits_{\alpha =0}^{N+3} {\left. {{{\partial x^\mu }
\over {\partial Z^\alpha }}} \right|_{\varsigma =0}\Gamma ^\alpha }.\eqno(4.8)
$$
 The Fourier decomposition of the spinor fields with respect to the arguments
$\varsigma ^a$ is
$$
\Phi '(x^\mu ,\varsigma ^a)=l^{-N/2}\sum\limits_n^{} {\Phi (x^\mu ,n)}\exp
(i2\pi n^a\varsigma ^a/l).\eqno(4.9)
$$
 Substituting the above equation into the action (4.6) and performing the
$\varsigma ^a$ integrations, we obtain the action of four dimensional space
$$
\eqalign{&S_{spinor}[\Phi ,X]\cr
  &=\int\limits_{\  M}^{} {\sqrt {-g(x)}dx^4\left( {{{i\bar \Phi (x,n)} \over
2}\Gamma ^\mu (x){{\partial \Phi (x,n)} \over {\partial x^\mu }}} \right.}\cr
  &\  -{{i\partial \bar \Phi (x,n)} \over {2\partial x^\mu }}\Gamma ^\mu
(x)\Phi (x,n)+{{i2\pi n^a} \over \ell }\bar \Phi (x,n)\left. {\Gamma ^\alpha
{{\partial \varsigma ^a} \over {\partial Z^\alpha }}} \right|_{\varsigma
=0}\Phi (x,n)\left. {} \right),\cr}\eqno(4.10)
  $$
 Although $\Phi (Z^\alpha )$ transforms as a spinor under the Lorentz
transformation of $(Z^\alpha )$, $\Phi '(x^\mu ,\varsigma ^a)$
$=\Phi (Z^\alpha (x^\mu ,\varsigma ^a)),$ is a set of scalar fields with
respect to the coordinates $x^\mu $. Therefore we have to perform a local
transformation on the fields $\Phi (x,n)$to make fields transforming as spinors
on $M$. The transformation is given as follows
$$
{\Phi (x,n)\to \Psi (x,n)=\rho ^{-1}(e(x))\Phi (x,n)},\eqno(4.11)
$$
where $e(x)$ is an SO(1,N+3) matrix field on $M$ and $\rho $ is a spinor
representation of $e(x)$. The field  $e(x)$ is defined by putting N+4 column
vectors\footnote*{ In the language of differential geometry, the set of these
ordered $N+4$ vectors is called an $B%a(Jadapted frame$B%b(J.} $e_i(x)$
%%(i=0,..., N+3) on $M$ successively
$$
e(x)=\left[ {e_0(x)...e_3(x) \ e_4(x)...e_{3+N}(x)} \right]. \eqno(4.12)
$$
 Here $e_0(x),...,e_3(x)$ are orthonormal vectors (in $V$) tangent to $M$ and
$e_4(x),$$...,$
$e_{3+N}(x)$ are othonormal vectors normal to $M$ [5]. The action (4.10) is
written in terms of $\Psi (x,n)$
$$
\eqalign{&S_{spinor}[\Psi ,X]=\int\limits_M {\sqrt {-g(x)}dx^4\ \sum\limits_n
{i\bar \Psi (x,n)[{1 \over 2}(\gamma ^\mu \vec \nabla _\mu
-\mathord{\buildrel{\lower3pt\hbox{$\scriptscriptstyle\leftarrow$}}\over \nabla
} _\mu \gamma ^\mu )+iM_n}}\cr
  &+i{1 \over 2}(\gamma ^\mu \phi _\mu +\phi _\mu \gamma ^\mu )]\Psi
(x,n),\cr}\eqno(4.13)
 $$
where
$$
\vec \nabla _\mu \equiv {{\vec \partial } \over {\partial x^\mu }}+\omega _\mu
^{lm}(e(x)){{[\gamma _l,\gamma _m]} \over 8}+A_\mu ^{ab}(e(x)){{[\hat \Gamma
_a,\hat \Gamma _b]} \over 8},\eqno(4.14a)$$
 $$\mathord{\buildrel{\lower3pt\hbox{$\scriptscriptstyle\leftarrow$}}\over
\nabla } _\mu \equiv
{{\mathord{\buildrel{\lower3pt\hbox{$\scriptscriptstyle\leftarrow$}}\over
\partial } } \over {\partial x^\mu }}-\omega _\mu ^{lm}(e(x)){{[\gamma
_l,\gamma _m]} \over 8}-A_\mu ^{ab}(e(x)){{[\hat \Gamma _a,\hat \Gamma _b]}
\over 8},\eqno(4.14b)
$$
$$
\phi _\mu (x)=\sum\limits_{l=0}^3 {\sum\limits_{a=1}^N {\phi _\mu ^{l\
3+a}{{[\gamma _l,\Gamma _{3+a}]} \over 4}}},\eqno(4.14c)$$
$$
\gamma ^\mu (x)=\sum\limits_{n=0}^3 {e^\mu _n(x)\gamma ^n}.\eqno(4.15)$$
$$
M_n={{2\pi n^a\Gamma ^{3+a}} \over \ell }\eqno(4.16)
$$
$e^\mu _n(x)$( n=0,1,2,3) are vierbeins and $\omega _\mu $,  $A_\mu $ and $\phi
_\mu ^{l\  3+a}$ are defined in terms of $e(x)$,
$$
\omega _{\mu \   m}^{\   l}(e(x))=\sum\limits_{\alpha =0}^{N+3}
{[e^{-1}(x)]_\alpha ^l}{{\partial [e(x)]^\alpha _m} \over {\partial x^\mu
}},\eqno(4.17)$$
$$
A_{\mu \  b}^{\  a}(e(x))=\sum\limits_{\alpha =0}^{N+3} {[e^{-1}(x)]_\alpha
^{a+3}}{{\partial [e(x)]^\alpha _{b+3}} \over {\partial x^\mu }},\eqno(4.18)$$
$$
\phi _\mu ^{\  \l \ 3+a}=\sum\limits_{\alpha =0}^{N+3} {[e^{-1}(x)]_\alpha
^l{{\partial [e(x)]^{\alpha 3+a}} \over {\partial x^\mu }}}.\eqno(4.19)
$$
 To prove the equation (4.13) we used the following relations
$$
\rho ^\dagger (e(x))\Gamma ^0=\Gamma ^0\rho ^{-1}(e(x)),\eqno(4.20)
$$
$$
\rho ^{-1}(e(x))\Gamma ^\mu \rho (e(x))=\gamma ^\mu (x),\eqno(4.21)$$
$$
\eqalign{&\rho ^{-1}(e(x)){{\partial \rho (e(x))} \over {\partial x^\mu }}\cr
  &={1 \over 2}\sum\limits_{l,m=0}^3 {\omega _\mu ^{lm}\rho _*(T^{lm})}+{1
\over 2}\sum\limits_{a,b=1}^N {A_\mu ^{ab}\rho _*(T^{3+a\
3+b})}+\sum\limits_{l=0}^3 {\sum\limits_{a=1}^3 {\phi _\mu ^{la}\rho _*(T^{l\
3+a})}},\cr}\eqno(4.22)
  $$
where $\rho _*$ is a differential representation [8] of $\rho $ and $T^{ij}$ (
i,j=0,1,...,N+3 ) are the generators of SO(1,N+3). We find that the last term
in the action (4.13) vanishes, using the symmetric property of the extrinsic
curvature ( the second fundamental tensor ) $\phi _{\mu \nu }^{3+a}$
$$
\phi _{\mu \nu }^{3+a}=\phi _{\nu \mu }^{3+a}.\eqno(4.23)
$$
 After integrating by part, we find that eq.(4.13) is written as
$$
S_{spinor}[\Phi ,X]=\int\limits_M^{} {\sqrt {-g(x)}dx^4\ i\sum\limits_n^{}
{\bar \Psi (x,n){\gamma ^\mu (x)\vec \nabla _\mu }+iM_n}\Psi (x,n)}.\eqno(4.24)
$$
 We can choose another N+4 vectors as
$$
e'_l(x)=\sum\limits_{m=0}^3 {e_m(x)\Omega ^m_l(x)}\  (l=0,1,2,3),\eqno(4.25)
$$
$$
e'_{a+3}(x)=\sum\limits_{b=1}^N {e_{b+3}(x)U^b_a(x)}\
(a=1,2,...,N),\eqno(4.26)
$$
where $\Omega (x)$ and $U(x)$ belong to $SO(1,3)$ and $SO(N)$ respectively. The
fields $\omega _\mu $transform as spin connections under (4.25) and $A_\mu $
transform as SO(N) gauge fields under (4.26) [5]. It should be noted that SO(N)
gauge fields thus induced are not flat in general and that the gravity and
SO(N) gauge fields  are not independent any more. In the language of
differential geometry, the fields $\omega _\mu $and $A_\mu $ correspond to
Levi-Chivita connection and normal connections of $M$ respectively. Furthermore
$\Psi (x,n)$ transform as $2^{[N/2]}$ plet of Dirac spinor fields on $M$ under
(4.25) and (4.26). Therefore we can identify the transformation (4.25) as a
local Lorentz transformation and (4.26) as an SO(N) gauge transformation. The
matrices $iM_n$ behave as mass matrices of the spinor fields $\Psi (x,n)$.
 Finally we comment on the focal length of the embedded submanifold introduced
in  section three. Following Milnor[9], the eigenvalues $K_i^{(a)}$(i=0,1,2,3)
of the matrices $[K^{(a)}]_m^l=\phi _m^{l\  a}$(a=1,...,N) are called principal
curvatures of $M$ in the normal direction $e_{3+a}(x)$. The reciprocals
$K_i^{(a)-1}$ (i=0,1,2,3) of these principal curvatures are called  principal
radii of the curvatures. It is known that the focal points\footnote*{ The focal
points exist at most four in this direction.}of $p\in M$, in the direction
$e_{3+a}(x)$, are determined by the equations
$$
Z^{(i)\alpha }=X^\alpha (x)+K_i^{(a)-1}[e_{3+a}(x)]^\alpha \
(i=0,1,2,3).\eqno(4.27)
$$
 Here $Z^{(i)\alpha }$ are coordinates of the focal points and $X^\alpha $ are
coordinates of $p\in M$.  The focal lengths of these focal points are given by
$K_i^{(a)-1}$.

\beginsection 5 Trace of extrinsic curvature as Higgs fields

In the previous section, we found that the SO(N) gauge fields and the gravity
appear\footnote{**}{A unified aspect of gravity and gauge fields by using a
solitonic solution in higher dimensional space was discussed by Akama [10].} in
the action (4.24). We saw, however, that the extrinsic curvature did not remain
in the final expression. We would hope that the extrinsic curvature would be
responsible for spontaneous symmetry breaking of gauge symmetry. In this
section we show that an appropriate surface term added to the action (4.5)
induces a Yukawa interaction term and that the trace of the extrinsic curvature
plays a role as Higgs fields.
 We consider the following surface term added to the action (4.5),
$$
S'=\int_{V_*} {dZ^{N+4}}k\bar \Phi \Gamma ^\alpha ({{\vec \partial } \over
{\partial Z^\alpha
}}+{{\mathord{\buildrel{\lower3pt\hbox{$\scriptscriptstyle\leftarrow$}}\over
\partial } } \over {\partial Z^\alpha }})\Phi .\eqno(5.1)
$$
 Following the same procedure as in the previous section, we find
$$
\eqalign{&S'=\int\limits_M {dx^4\sqrt {-g(x)}}k\sum\limits_n {[\bar \Psi
(x,n)}\gamma ^\mu (x)\nabla _\mu \Psi (x,n)+\nabla _\mu \bar \Psi (x,n)\gamma
^\mu (x)\Psi (x,n)\cr
  &+\bar \Psi (x,n)(\gamma ^\mu (x)\phi _\mu -\phi _\mu \gamma ^\mu (x))\Psi
(x,n)].\cr
  &\cr}\eqno(5.2)
  $$
 After integrating by parts, we obtain
$$
S'=\int\limits_M {dx^4\sqrt {-g(x)}}k\bar \Psi (x,n)\not \phi \Psi
(x,n).\eqno(5.3)
$$
where
$$
\not \phi =\phi ^a\Gamma ^{3+a},\eqno(5.4)$$
$$
\phi ^a\equiv \sum\limits_{l=0}^3 {\phi _l^{l\  3+a}}.\eqno(5.5)
$$
 The fields $\phi ^a$ ( a=1,...,N ) transform as an N-plet of scalar fields. In
the next paper, we shall show that these scalar fields behave as Higgs fields.

 \beginsection 6 Discussion

 We discussed the classical dynamics of free fields contained in a neighborhood
of a four dimensional submanifold in $V$. The classical dynamics of the systems
is described by an infinite number of fields in the four dimensional
submanifold interacting with gravity, SO(N) gauge fields and an N-plet of
scalar fields in the limit
 ;$l/d(X)\to 0$.( The gravity, SO(N) gauge fields and N-plet of scalar fields
are not independent however. ) When $l$ is very small; the mass of $n$($\ne 0$)
modes are very large, the  massive modes  cannot be excited classically in low
energy physics. Only the zero modes remain in the eq. (3.11), (4.24) and (5.3).
We find, however, that quantum effects caused by the massive modes are not
negligible. The massive modes of the fields are influenced by the configuration
of the submanifold, which itself provides the gravity, the SO(N) gauge fields
and the N-plet of the scalar fields.
 It is therefore naturally expected that the quantum effects caused by the
massive modes generate dynamics of the gravity, the SO(N) gauge fields and the
N-plet of the scalar fields. In the next paper, we will show that Einstein
SO(N)-Yang-Mills-Higgs theory is induced as a low energy effective theory of
the system and that $\phi ^a$ fields\footnote*{The fields $\phi$ have non zero
vacuum expectation values and break $SO(N)$ gauge invariance spontaneously and
generate mass of gauge fields and fermions.} behave as Higgs fields. Such an
idea was first introduced by Saharov and Zel$B%f(Jdovich in the gravitational
%%case[11][12][13].
  The theory we developed here is different from Kaluza-Klein theory in many
respects. In Kaluza-Klein type theory, the higher dimensional space is curved
and has its own dynamics. Our embedding space $V$ is flat and dose not change.
In Kaluza-Klein theory, extra dimensions of the higher dimensional space must
be compactified to make four dimensional spacetime. In our theory , spacetime
is embedded into $V$. See Fig[4].

 In our theory, we do not have to assume compactifications of higher
dimensional space. Instead we assume the existence of the
$B%a(Jtemplate$B%b(J of spacetime which confines the fields inside the
%%neighborhood of the arbitrarily embedded four dimensional submanifold.

\beginsection ACKNOWLEDGMENTS

 I am very grateful to K. Kikkawa,H. Itoyama, H. Kunitomo, K. Shizuya and H.
Suzuki and  for useful discussions and comments.

\beginsection REFERENCES

\item{[1]} H.B.G Casimir, Proc. K. Ned. Acad. Wet. 51 (1948),793

\item{[2]} J. Ambjorn and S. Wolfram, Ann. Phys. 147 (1983), 1

\item{[3]} R. Balian and B. Dulplantie, Ann. Phys. 112 (1978),165
      K. A. Milton, L. L. De Raad, Jr., and J. Schwinger, Ann. Phys. 115
(1978), 388

\item{[4]} T. Isse,   (to appear )

\item{[5]} S.Kobayashi. and K.Nomizu, Foundation of Differential Geometry  vol
I, II ( Interscience Publishers , New York, 1969)

\item{[6]} Th. Kaluza, sitzungsber. Preuss. Akad. Wiss. Phys. Math. K1. 966
(1926)

\item{[7]} T.Appelquist et al., eds., Modern Kaluza-Klein Theory
(Addison-Wesley , New York, 1987)

\item{[8]} R. Hermann, Lie Groups for Physicists ( W.A. Benjamin. INC, Menlo
Park, 1966)

\item{[9]} J. Milnor, Morse Theory ( Princeton Univ. Press , Princeton, 1963 )

\item{[10]} K. Akama, Prog. Theor. Phys. 78 (1987), 184

\item{[11]} A. D. Saharov, Sov. Phys. Doklady 12 (1968), 1040

\item{[12]} Ya. B Zel$B%f(Jdovich, Zh.  Eksp. Teor. Fiz. Pis$B%f(Jma Red 6
%%(1967), 883

\item{[13]} for review see, S.L. Adller, Rev. Mod. Phys., 54 (1982), 729

 \end